# Metaverse Shape of Your Life for Future: A bibliometric snapshot

Muhammet Damar
*Dokuz Eylül Universty*
İzmir, Turkey
muhammet.damar@deu.edu.tr
0000-0002-3985-3073

*Abstract*— The metaverse was first introduced in 1992. Many people saw Metaverse as a new word but the concept of Metaverse is not a new term. However, Zuckerberg's press release drew all the attention to the Metaverse. This study presents a bibliometric evaluation of metaverse technology, which has been discussed in the literature since the nineties. A field study is carried out especially for the metaverse, which is a new and trendy subject. In this way, descriptive information is presented on journals, institutions, prominent researchers, and countries in the field, as well as extra evaluation on the prominent topics in the field and researchers with heavy citations. In our study, which was carried out by extracting the data of all documents between the years 1990-2021 from the Web of Science database, it was seen that there were few studies in the literature in the historical process for the metaverse, whose popularity has reached its peak in recent months. In addition, it is seen that the subject is handled intensively with virtual reality and augmented reality technologies, and the education sector and digital marketing fields show interest in the field. Metaverse will probably have entered many areas of our lives in the next 15-20 years, shape our lives by taking advantage of the opportunities of developing technology.

## I. INTRODUCTION

Mark Zuckerberg announced in October, 2021 that Facebook will change its name to Meta and also make significant investments in Metaverse [13]. Many people saw Metaverse as a new word. But the concept of Metaverse is not a new term. The word metaverse first appeared in 1992 in a speculative piece of fiction called Snow Crash by Neal Stephenson. In this novel, Stephenson defines the metaverse as a large virtual environment [9]. More recently, a metaverse called OASIS has been featured in the novel and movie Ready Player One [22]. In the novel, the metaverse is a virtual shared space that combines virtual reality, augmented reality, and the internet [13]. But in the intervening 30 years, the concept has never been more popular. Developments show that the concept will make itself mentioned even more.

Alang [1] defined the Metaverse as the layer between you and reality. Metaverse refers to a 3D virtual shared world where all activities can be carried out with the help of augmented and virtual reality services. Such platforms have gained popularity over the past few years as people shift their activities online, especially during the coronavirus pandemic [25]. The metaverse refers to a shared 3D virtual world in which all activities can take place using augmented and virtual reality equipment. These platforms have gained popularity in recent years as people shift their activities online, especially during the COVID-19 pandemic [6].

With the SARS-CoV-2 virus, which entered our lives in January 2020, and the devastating effect of this virus on our lives, humanity is deprived of many opportunities in its daily life. Hundreds of millions of people around the world have had our lives paralyzed by the spread of COVID-19, the viral disease of the coronavirus, and its associated severe acute respiratory syndrome. Due to the contagious effect of COVID-19 and the desire to prevent this transmission, humanity has been subject to restrictions in its movements [7,11]. This disease, which affects all sectors from the education sector to the production sector, has further increased the importance of concepts such as distance work or distance education. Masters et al. [12] stated that the importance of distance education has increased in the increasing Covid 19 disease, and the ways to make it more realistic in order to increase the effectiveness and range of education are being questioned.

Zuckerberg's vision didn't have to do with advertising, which provides the bulk of Facebook's current profits, or the increase in the total size of the social network, which already has around three billion monthly active users [2]. In summary, Facebook's introduction by Mark Zuckerberg has shifted everyone's attention to Metaverse and has made it a topic that is discussed in our daily lives. Humanity has begun to question the concept of the Metaverse even more now. At first glance, the concept of the metaverse is directly associated with augmented reality and virtual reality issues. But it is also clear that it is much more than what Mark Zuckerberg imagined or what the Seoul municipality of South Korea imagined.

As Zuckerberg revealed to the media before the Facebook Connect conference, the company claims it will be the next big computing platform after the rise of smartphones and the mobile web [8]. Not only Facebook's Meta company, but also companies such as Microsoft and Nike are struggling to take their place in this market. Even states are getting hungry for national policies on this issue, with South Korea the only government trying to recreate the virtual public square [23]. In summary, it is not just private businesses that support Metaverse. The South Korean State has started to carry out policies on this issue and they have started to share it with the public. The first statement on this issue came from South Korea. It has been announced that a large enterprise involving 500 companies including Samsung, Hyundai Motors, SK Telecom, and KT will be provided by the state, for example, $26 million in just 2022 [10]. The future of the Metaverse is almost entirely built by companies.

Alang [1] stated that the first thing to ask about the metaverse is whether it is something wanted. He also stated that a great deal of work will be done in the process of providing technological development for the metaverse and that this is not a single product presentation, but much more. For this, he gave the example of the Uber initiative and stated that Uber is not just a car calling application on your phone, but actually, an infrastructure that includes the entire transportation infrastructure. He stated that the company aims to be the technology layer that not only drives rides but also drives bicycles, scooters, public transport, and ultimately autonomous cars. But Uber sold its autonomous car division after it became clear that solving self-driving is a much more difficult problem than anyone knows.

Shotton et al.[19] stated that the ability to explore, integrate and reuse relevant scientific outputs from previous studies is critical for innovative research [19]. He emphasized the importance of information regulation to shape the future of societies and stated that information has become the most important resource for the future of late modern societies [17]. With the processing of information in the right hands, a strategy can emerge for institutions, countries, and businesses. Rotolo et al. [15] said that Conventional tools of strategic intelligence include science and technology foresight, innovation policy evaluation, and technology assessment. Thanks to the fields of bibliometry or scientometrics, it can be used for common science and technology assessment to see the outputs of policies and activities carried out in certain periods in such strategic associations. Metaverse is one of the most up-to-date technological developments in the rapidly transforming world in this regard. It is valuable to observe and follow the equivalent of the discussions in this direction in the literature.

Zuckerberg betting that it will be the next big computing platform for the metaverse, after the rise of smartphones and the mobile web [8]. Will this expectation come true, as in the case of Uber, and what kind of discussions have been made in the literature in this direction from past to present. At this point, the bibliometric field provides good tools for exploring the texture of work done in a particular field or topic. Bibliometric studies are important studies to define a particular field and to better understand the researcher's texture in that field. In this way, researchers working in the field can have information and ideas about the development of the literature and researchers who have come to the fore in the field. Likewise, there are important journals that stand out in certain fields. The motivation for this study is the lack of a bibliometric study in the literature on a topic that has been so popular recently. It is thought that it will be a guide for researchers and will provide information about important congresses, journals, and institutions of the field.

## II. METHOD

This study, publications in the field of "metaverse", which are in the Web of Science database and scanned by SCI-EXPANDED, SSCI, A&HCI, CPCI-S, CPCI-SSH, ESCI indexes, were evaluated. Data were collected on 26.11.2021. The search keywords used in the study is as follows: (ti="metaverse" or ak="metaverse" or kp="metaverse")

93 documents were accessed in the Web of Science database, 1560 different sources and a total of 1855 different references were used in 93 documents. It has been seen that related studies have been carried out by 155 different authors, 29 different countries, and 96 different organizations. The research questions that the study focuses on are as follows:

- What are the distribution, languages and document types of documents produced on Metaverse by years?
- What is the situation of researchers, institutions and countries doing intensive research on the Metaverse?
- What is the bibliometric trends according to published documents types such as book chapters, proceedings and articles on the Metaverse?
- What are the journals that publish the most on Metaverse, the publishers that come to the fore, and the congresses that stand out in the field?
- What are the most frequently cited works on Metaverse?
- What are the generally accepted debates on the Metaverse and what is the general trends of the topics discussed in the historical process from past to present?

Vosviewer program was used for bibliometric analysis in the study. VOSviewer is the software package for analysing and visualising large bibliographic datasets. VOSViewer applies its own algorithm that is a modularity-based clustering technique, which is similar to the multidimensional scaling and is based on the smart local moving algorithm [26,27].

## III. DISCUSSION AND FINDINGS

The findings obtained from the bibliometric analysis carried out in the study will be carried out in the form of general view, focus on book chapters, focus on proceedings, focus on the article, and finally content analysis for all documents. These contents are presented under the headings in order below.

### A. General View

The number of documents accessed via Web of Science is 93. These publications in total times cited:281 and average per item:3.02 and our h-index value is 9. These documents were produced by 155 different researchers. When sorted these documents according to their types; articles (f:42), proceedings papers (f:38), book chapters (f:14). In the Web of Science environment, a document can be included in the same class as both articles and book chapters as a document type.

When evaluated document productivity by years; 2021(f:6), 2020 (f:5),2019 (f:3), 2018 (f:5), 2017 (f:5), 2016 (f:6),2015 (f:12), 2014 (f:2),2013 (f:6),2012 (f:5), 2011 (f:6),2010 (f:10),2009 (f:9), 2008 (f:7), 2007 (f:2), 2006 (f:1), 2000 (f:1), 1996 (f:1), 1995 (f:1). Publishing organizations that have produced a minimum of two documents when sorted according to Publishers; Springer Nature (f:20), IEEE (f:15), IGI Global (f:7), Elsevier (f:6), Assoc Computing Machinery (f:5), MDPI (f:3), Kassel Univ Press Gmbh (f:2), Sage (f:2), Taylor & Francis (f:2).

When ranked the obtained works according to the researach area; computer science (f:42), engineering (f:17), education educational research (f:8), psychology (f:8), art (f:7), business economics (f:7), information science library science (f:5), science technology other topics (f:5), communication (f:4), cultural studies (f:4), imaging science photographic technology (f:3), public administration (f:3), religion (f:3), telecommunications (f:3), arts humanities other

*Journal of Metaverse*

*Damar, M.*

topics (f:2), chemistry (f:2), geography (f:2), philosophy (f:2), environmental sciences ecology (f: 1), government law (f:1), health care sciences services (f:1), instruments instrumentation (f:1), music (f:1), social issues (f:1), social sciences other topics (f :1), sociology (f:1).

When the studies are evaluated according to the subject area, it is seen that they concentrate on computer science and engineering (f:17). However, when the research area texture is evaluated, it can be seen that the studies on the metaverse find application in many different fields, from the field of health to the field of education, from the field of culture to the field of environmental sciences. It is seen that metaverse studies are carried out on psychology, religion, philosophy, and social issues due to human-computer interaction.

When ranked the works according to the languages in which they were produced; English (f:87), Spanish (f:4), Portuguese (f:1), Russian (f:1). The countries that come to the fore in this productivity are respectively; USA (f:19), Japan (f:10), Brazil (f:8), South Korea (f:8), England (f:7), Turkey (f:7), Spain (f:5), Belgium (f:3), Colombia (f:3), France (f:3), Sweden (f:3). The data of book chapters, proceedings, and articles are presented below, respectively. While South Korea has come to the forefront in studies and the expansion of countries, South Korea is behind the USA and Japan in terms of the metaverse. In addition, although recently, both Facebook manager Zuckerberg's statement [2,8,10], it has been seen that the subject that have discussed extensively in daily life has been evaluated by few studies and researchers in the literature. It is thought that the literature on metaverse will be enriched after the concept is overlapped with the words augmented reality and virtual reality, and there is more work in the literature in this direction, the definition of a general concept as an umbrella term as metaverse, and the discourses of important companies and some states are developing policies.

*B. Focus on Book Chapters*

Distribution of book chapter studies on Metaverse by years; 2017 (f:1), 2015 (f:7), 2013 (f:3), 2012 (f:1), 2010 (f:1), 2009 (f:1). A total of 14 book chapter studies were carried out and related studies received 25 citations in total. All of the book section studies have been published in the English Language. Book chapters' publishers are IGI Global (f:7), Springer Nature (f:6), Ashgate Publishing Ltd (f:1). Book Series Titles are Advances In Educational Technologies And Instructional Design Book Series (f:7), Human-Computer Interaction Series (f:1), New Directions In Planning Theory (f:1), Smart Computing and Intelligence (f:1).

Researchers who carried out the work of the book section; Backes L (f:7), Schlemmer E (f:7), Calongne C (f:1), Dede CJ (f:1), Devisch O (f:1), Huvila I (f:1), Jacobson J (f:1), Lombardi J (f:1), Lombardi M (f:1), Power D (f:1), Richards J (f:1), Sheehy P (f:1), Sonvilla-weiss S (f:1), Sonvillaweiss S (f:1), Stricker A (f:1), Teigland R (f:1). Distribution of researchers who made book chapters to countries; Brazil (f:7), USA (f:3), Sweden (f:2), Belgium (f:1), Finland (f:1), Norway (f:1). It was seen that Brazil came to the fore in the book chapters. The reason for this issue is the study of the book Learning in Metaverses: Co-Existing in Real Virtuality: Co-Existing in Real Virtuality [16], edited by Eliane Schlemmer in the Metaverse field in the relevant country.

The institutions of the researchers who carried out the book section are as follows; Centro Universitario La Salle (f:7), Universidade Do Vale Do Rio Dos Sinos Unisinos (f:7), Uppsala University (f:2), Aalto University (f:1), Colorado Technology University (f:1), Consulting Services for Education (f:1), Duke University (f:1), Enterprise VR (f:1), Harvard University (f:1), PHL University College in Belgium (f:1), Stockholm School of Economics (f:1), University of Agder (f:1). In addition, the distribution of book chapter studies according to research areas is as follows: psychology (f:8), communication (f:3), cultural studies (f: 3), computer science (f:2), art (f:1), education educational (f:1), geography (f:1), public administration (f:1).

*C. Focus on Proceedings*

The progress of the papers by years is as follows: 2019 (f:2), 2018 (f:3), 2016 (f:4), 2015 (f:4), 2014 (f:2), 2013 (f:2), 2012 (f:3), 2011 (f:4), 2010 (f:5), 2009 (f:4), 20082), 2007 (f:1), 2006 (f:1), 2000 (f:1). When this course is evaluated, there is no significant jump. It has been seen that 3-4 papers are presented in the related congresses a year. Facebook and South Korea's serious releases on the metaverse and their sharing of their strategic action plans with the public promises hope in the number of papers produced in the coming years and even in the planning of new congresses related to the metaverse.

When the papers produced on Metaverse are evaluated; Japan (f:10), USA (f:7), Turkey (f:4), England (f:3), Austria (f:2), France (f:2), Netherlands (f:2), South Korea (f:2), Spain (f:2). When the papers are evaluated according to the institutions where they are produced (minimum 2), the following table is encountered; Sabanci University (f:4), Clarkson University (f:3), Ritsumeikan University (f:3), Suzuka College (f:3), Gifu College (f:2 Graz University of Technology (f:2), Nagaoka University of Technology (f:2), Tsuyama College (f:2), University of La Reunion (f:2) There are not very serious differences between countries and related countries.

The prominent researchers in the paper are as follows. Ayiter E (f:4), Barry DM (f:4), Fukumura Y (f:4), Kanematsu H (f:4), Ogawa N (f:4 Kobayashi T (f:3), Thawonmas R (f) :3), Conruyt N (f:2), Dharmawansa A (f:2), Sebastien D (f:2), Shirai T (f:2), Yajima K (f:2) When both institutions and researchers are evaluated, It is thought that the relevant data will be a reference for researchers who want to do post-doctoral research on the metaverse, who want to do their postgraduate education in the metaverse field, and who are looking for an institution.

When the papers are sorted according to the research areas in which they are located, the following view is encountered; computer science (f:35), engineering (f:12), imaging science photographic technology (f:3), business economics (f:2), education educational research (f:2), telecommunications (f:2), chemistry (f:1), health care sciences services (f:1), information science library science (f:1), philosophy (f:1), public administration (f:1), religion (f:1), science technology other topics (f:1), social issues (f:1). It can be stated that the trend in the field of proceedings research is similar to the general trend.

The conferences that come to the fore in the field for papers can be expressed as follows; International Conference on Cyberworlds, International Conference on Entertainment Computing, Biennial Pan Ocean Remote Sensing Conference Porsec, International Conference on Entertainment Computing, International Conference on E-Commerce and

*Journal of Metaverse*

*Damar M.*

Web Technologies, World Marketing Congress on Looking Forward Looking Back Drawing on the Past to Shape the Future of Marketing, International Conference on Knowledge-Based and Intelligent Information and Engineering Systems, IEEE International Conference on Games and Virtual Worlds for Serious Applications vs Games, International Conference on Advanced Information Networking and Applications, International Conference on Virtual Reality Held at the Human-Computer Interaction , International Conference of the Chilean Computer Science Society, European Conference on Games Based Learning, International Conference on Intelligent Interactive Multimedia Systems and Services, International Conference on Internet and Web Applications and Services, International Technology Education and Development Conference, International Conference on Intelligent Games and Simulation, Augmented Human International Conference, Virtual Reality International Conference, Conference on Engineering Reality of Virtual Reality. Related congresses can also be expressed as congresses where the most intense discussions about its development in the field are held. It can be expressed as qualified congresses where researchers can present their scientific studies in the field.

TABLE I. TOP TEN PROCEEDINGS ABOUT METAVERSE

| Rank | Title | Year | C |
|---|---|---|---|
| 1 | Making real money in virtual worlds: MMORPGs and emerging business opportunities, challenges and ethical implications in metaverses | 2008 | 153 |
| 2 | 3D Virtual Worlds and the Metaverse: Current Status and Future Possibilities | 2013 | 111 |
| 3 | Introduction: Virtual, Augmented, and Mixed Realities in Education | 2017 | 85 |
| 4 | User-Friendly Home Automation Based on 3D Virtual World | 2010 | 70 |
| 5 | Retail spatial evolution: paving the way from traditional to metaverse retailing | 2009 | 62 |
| 6 | A content service deployment plan for metaverse museum exhibitions-Centering on the combination of beacons and HMDs | 2017 | 53 |
| 7 | Retailing in Social Virtual Worlds: Developing a Typology of Virtual Store Atmospherics | 2015 | 28 |
| 8 | Virtual world, defined from a technological perspective and applied to video games, mixed reality, and the Metaverse | 2018 | 21 |
| 9 | Metaverses as a Platform for Game Based Learning | 2010 | 19 |
| 10 | Innovation and imitation effects in Metaverse service adoption | 2011 | 19 |
| 11 | Synthetic Educational Environment - a Footpace to New Education | 2017 | 18 |
| 12 | Multilingual Discussion in Metaverse among Students from the USA, Korea and Japan | 2010 | 15 |
| 13 | Splendid isolation: 'Philosopher's islands' and the reimagination of space | 2012 | 14 |
| 14 | Virtual STEM class for nuclear safety education in metaverse | 2014 | 11 |
| 15 | Opening the Metaverse | 2010 | 15 |
| 16 | Distributed Metaverse: Creating Decentralized Blockchain-based Model for Peer-to-peer Sharing of Virtual Spaces for Mixed Reality Applications | 2018 | 10 |
| 17 | Evaluation For Students' Learning Manner Using Eye Blinking System in Metaverse | 2015 | 10 |
| 18 | Blinking Eyes Behaviors and Face Temperatures of Students in YouTube Lessons - For the Future E-learning Class | 2016 | 8 |
| 19 | Virtual World as a Resource for Hybrid Education | 2020 | 7 |
| 20 | From Industry 4.0 to Nature 4.0-Sustainable Infrastructure Evolution by Design | 2018 | 6 |

[a.] Citations: C

When the general titles of the congresses related to the metaverse subject were evaluated, it was seen that the metaverse subjects were discussed more in the most intense computer science, information systems and engineer, intelligent systems congresses. However, along with these fields, the metaverse is also discussed intensively in congresses on e-health, education, e-commerce, web technology and digital marketing, interactive multimedia systems and services. In addition to these, it can be stated that the metaverse matures in these congresses and these congresses are important for its structuring, especially considering the 11th-12th conferences on augmented reality and human interaction, and the fact that the field will be built on virtual reality and augmented technologies.

Table 1 below presents the most frequently cited proceedings on the metaverse. The fact that the congresses in which the works are presented are not similar, the fact that almost 10 papers are presented in different congresses shows that the word metaverse does not find its exact equivalent in the literature and academic environment. In general, it was seen that the discussions went over topics such as augmented reality, virtual reality, intelligent interactive multimedia systems.

*D. Focus on Articles*

There are 42 articles produced on the subject of Metaverse, and these articles are 217 times cited, average per item 5,17 and h-index:7 in total. 42 articles were produced in the following languages; English (f:37), Spanish (f:4), Russian (f:1). The productivity frequency of the articles by years is as follows: 2021 (f:4), 2020 (f:5), 2019 (f:1), 2018 (f:2), 2017 (f:4), 2016 (f:2), 2015 (f:8), 2013 (f:3), 2012 (f:2),2011 (f:2), 2010 (f:4), 2009 (f:1), 2008 (f:2), 2000 (f:1), 1995 (f:1). As in the papers and book chapters, the development of the field in the literature and the interest in the word metaverse have been realized in a small number of articles, generally changing between 3-5 articles over the years. It can be stated that this situation is perhaps because the concept does not approach until 2021, like Mark Zuckerberg. Of the articles produced, 10 are scanned in the Emerging Sources Citation Index (ESCI), 12 in the Social Sciences Citation Index (SSCI), 9 in the Science Citation Index Expanded (SCI-EXPANDED), and 4 in the Arts & Humanities Citation Index (A&HCI). Others were scanned by Book Citation Index - Social Sciences & Humanities (BKCI-SSH), Book Citation Index - Science (BKCI-S), and Conference Proceedings Citation Index - Science (CPCI-S).

When the studies are evaluated according to the research area, they are as follows: psychology (f:8), computer science (f:6), business economics (f:5), education educational research (f:5), engineering (f:4), cultural studies ( f:3), art (f:2), arts humanities other topics (f:2), communication 2 geography (f:2), information science library science (f:2), public administration (f:2), religion (f:2), science technology other topics (f:2), chemistry (f:1), environmental sciences ecology (f:1), instruments instrumentation (f:1), music (f:1), philosophy (f:1), social sciences other topics (f:1), telecommunications (f:1).

The researchers and their institutions that have published the most intensive articles on the metaverse are given in Table 2. Backes L and Schlemmer E are Brazilian researchers who have come to the fore in the field with seven articles. When the researchers who carried out the studies are evaluated according to the countries; USA (f:10), Brazil (f:7), South Korea (f:6), England (f:4), Colombia (f:3), Spain (f:3), Turkey (f:3), Belgium (f:2), Sweden (f:2), Ecuador (f:1), France (f:1), Israel (f:1), Mexico (f:1), Peru (f:1), Ukraine (f:1).

*Damar, M.*

Fifteen journals and their research domains that stand out in the field are presented in Table 3. When the research domain of the journals is evaluated, computer science, software engineering, business; management, engineering, electrical & electronic seem to come to the fore.

TABLE II.  MOST PRODUCTIVE RESEARCHERS AND INSTITUTIONS PUBLISHING ARTICLES ABOUT METAVERSE

| Rank | Authors | Organizations | Country | Number of Doc. |
|---|---|---|---|---|
| 1 | Backes L | Centro Universitario La Salle | Brazil | 7 |
| 2 | Schlemmer E | Universidade do Vale do Rio dos Sinos | Brazil | 7 |
| 3 | Ayiter E | Sabanci University | Turkey | 2 |
| 4 | Bourlakis M | Cranfield University | England | 2 |
| 5 | Diaz JEM | Univ Cundinamarca | Brazil | 2 |
| 6 | Li F | City University London | England | 2 |
| 7 | Papagiannidis S | University of Newcastle | Australia | 2 |

a. Number of Document Count: N

TABLE III.  JOURNALS AND THEIR RESEARCH DOMAINS OF INTEREST IN METAVERSE-RELATED PUBLICATIONS

| Rank | Journal Title | JIF (2020) | Research Domain |
|---|---|---|---|
| 1 | ACM Computing Surveys | 14.098 | Computer Science, Theory & Methods |
| 2 | Computer Animation and Virtual Worlds | 1.020 | Computer Science, Software Engineering |
| 3 | Data Base for Advances in Information Systems | 1.828 | Information Science & Library Science |
| 4 | Electronic Commerce Research | 3.747 | Business; Management |
| 5 | Geoforum | 3.901 | Geography |
| 6 | Harvard Business Review | 6.870 | Business; Management |
| 7 | Human Centric Computing and Information Sciences | 5.900 | Computer Science, Information Systems |
| 8 | IEEE Technology and Society Magazine | 1.554 | Engineering, Electrical & Electronic |
| 9 | IEEE Transactions on Consumer Electronics | 2.947 | Engineering, Electrical & Electronic; Telecommunications |
| 10 | International Journal of Information Management | 14.098 | Information Science & Library Science |
| 11 | Journal of Consciousness Studies | 1.348 | Social Sciences, Interdisciplinary |
| 12 | Journal of Electronic Commerce Research | 2.861 | Business |
| 13 | Journal of Visual Culture | 0.400 | Cultural Studies |
| 14 | New Scientist | 0.319 | Multidisciplinary Sciences |
| 15 | Sensors | 3.576 | Chemistry, Analytical; Engineering, Electrical & Electronic; Instruments & Instrumentation |

b. Journal of Impact Factor: JIF

Table 4, The most prominent and most cited studies for researchers are as follows. In the study, the citation values obtained from Google Scholar were updated by taking the citation numbers from Web of Science as a reference.

*E. Content Analysis About Title, Abstract and Keywords*

In the analysis, the keywords most heavily associated with the word metaverse were second life, virtual worlds, avatar, 3d, augmented reality, virtual reality, virtual environments, art, mixed reality, collaboration, e-learning, multi-user virtual environment (muve), open simulator, virtualization, metaverse retailing, haptics, blockchain, industry 4.0 are the most frequently used keywords in published studies. In addition, words such as education, higher education, informal education, e-learning, schools, panoramic, museums, exhibition content, virtual excavation, church, smart maintenance, came to the fore. The reason for this is the intense work of virtual or augmented reality studies in educational studies. Boeing 737 maintenance education can be given as an example of the studies carried out in the field of education [21]. Recently, in the emerging world of mixed reality applications, various industries are already taking advantage of these technologies [4,5,18,20].

Metaverses embedded in our lives create virtual experiences according to part of life scenarios inside of the physical world (Siyaev & Jo, 2021). For example, the concept of haptics stands out among the researcher keywords used. Haptics differs from other robotic devices; It allows people to feel the shape, roughness, and vibration of the surface in a virtual reality environment or while controlling slave robots. Real perception is an important factor for the user to experience.

TABLE IV.  TOP 20 MOSTLY CITED ARTICLES ABOUT METAVERSE

| Rank | Title | Year | C |
|---|---|---|---|
| 1 | Making real money in virtual worlds: MMORPGs and emerging business opportunities, challenges and ethical implications in metaverses | 2008 | 153 |
| 2 | 3D Virtual Worlds and the Metaverse: Current Status and Future Possibilities | 2013 | 111 |
| 3 | Introduction: Virtual, Augmented, and Mixed Realities in Education | 2017 | 85 |
| 4 | User-Friendly Home Automation Based on 3D Virtual World | 2010 | 70 |
| 5 | Retail spatial evolution: paving the way from traditional to metaverse retailing | 2009 | 62 |
| 6 | A content service deployment plan for metaverse museum exhibitions-Centering on the combination of beacons and HMDs | 2017 | 53 |
| 7 | Retailing in Social Virtual Worlds: Developing a Typology of Virtual Store Atmospherics | 2015 | 28 |
| 8 | Virtual world, defined from a technological perspective and applied to video games, mixed reality, and the Metaverse | 2018 | 21 |
| 9 | Metaverses as a Platform for Game Based Learning | 2010 | 19 |
| 10 | Innovation and imitation effects in Metaverse service adoption | 2011 | 19 |
| 11 | Synthetic Educational Environment - a Footpace to New Education | 2017 | 18 |
| 12 | Multilingual Discussion in Metaverse among Students from the USA, Korea and Japan | 2010 | 15 |
| 13 | Splendid isolation: 'Philosopher's islands' and the reimagination of space | 2012 | 14 |
| 14 | Virtual STEM class for nuclear safety education in metaverse | 2014 | 11 |
| 15 | Opening the Metaverse | 2010 | 15 |
| 16 | Distributed Metaverse: Creating Decentralized Blockchain-based Model for Peer-to-peer Sharing of Virtual Spaces for Mixed Reality Applications | 2018 | 10 |
| 17 | Evaluation For Students' Learning Manner Using Eye Blinking System in Metaverse | 2015 | 10 |
| 18 | Blinking Eyes Behaviors and Face Temperatures of Students in YouTube Lessons - For the Future E-learning Class | 2016 | 8 |
| 19 | Virtual World as a Resource for Hybrid Education | 2020 | 7 |
| 20 | From Industry 4.0 to Nature 4.0-Sustainable Infrastructure Evolution by Design | 2018 | 6 |

c. Citations: C

Hardware and software are now sufficiently advanced. Metaverses that use virtual or augmented reality on affordable devices, pioneered by games, are likely to become much more popular. It can start to offer practical or fun features. And of course, there is a lot of money to be made. Fortnite generated $9 billion in revenue in 2018 and 2019 as people paid to customize their avatars [22]. This situation is also becoming appetizing for blockchain technology.

In Stephenson's metaverse, companies all pay for slices of digital real estate to an entity called the Global Multimedia Protocol Group. Users also pay for access; Only those who can afford cheaper common terminals appear as grainy black and white in the metaverse [2]. In fact, in the literature, it is seen that the subject has been discussed intensively, especially

## *Journal of Metaverse*

with blockchain technology, from the day the novel was written.

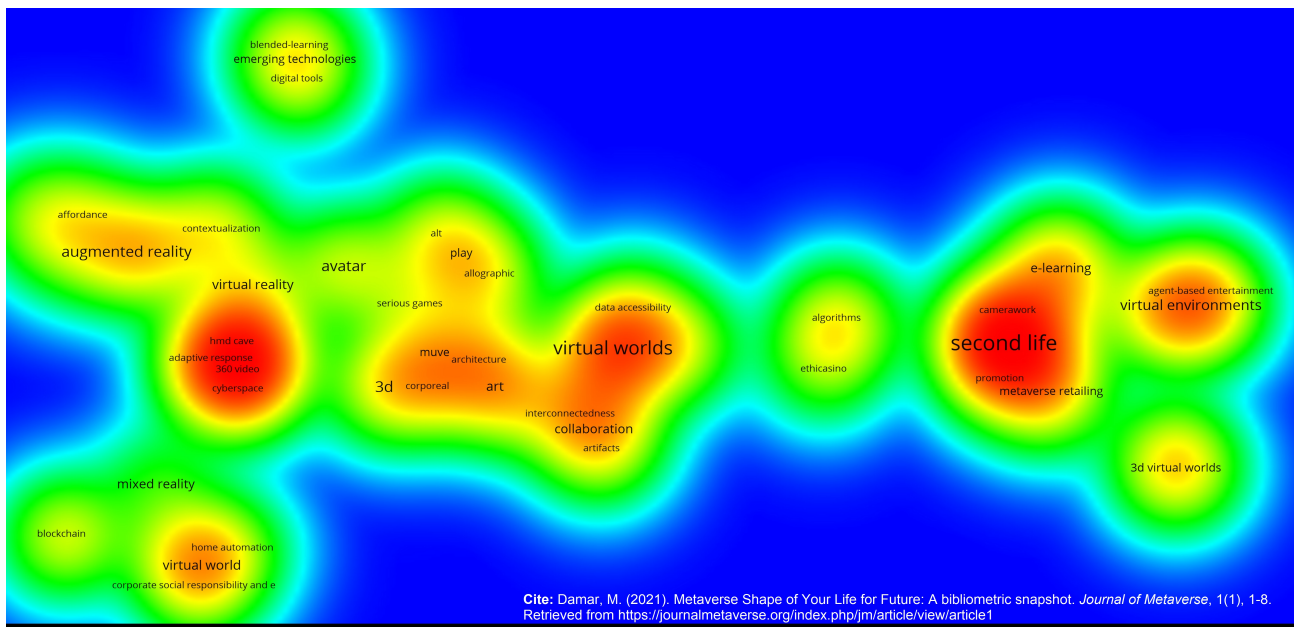

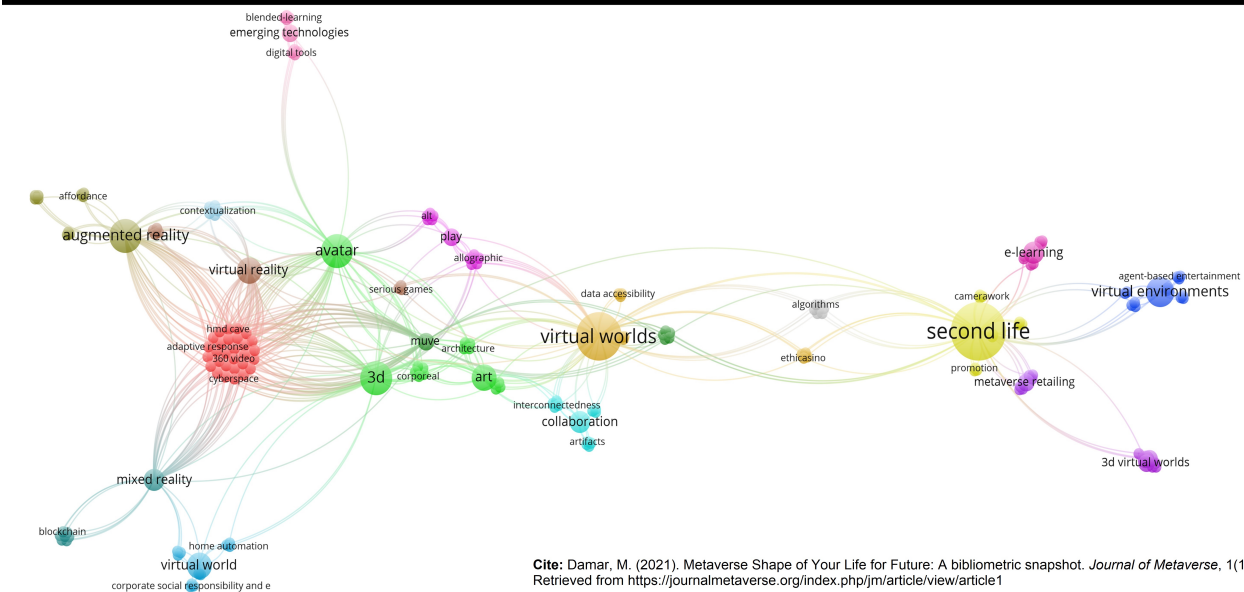

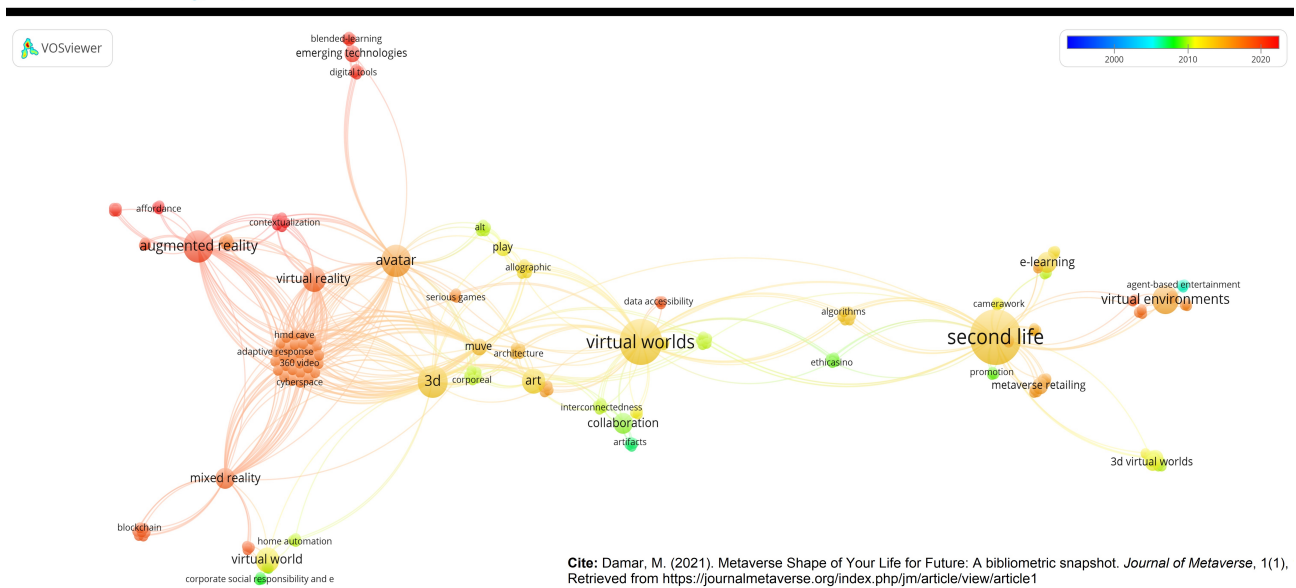

Fig. 1. Keyword Density, Overlay and Network Visualizations about Metaverse

The video game second life, released by Linden Lab in 2003, created a virtual world where users can navigate by building their structures [22]. In the literature, it has been seen that intensive studies have been made on Second Life. Roblox, a children's video game released in 2006, has recently evolved into an immersive world where players can design and sell their creations, from avatar costumes to their own interactive experiences [14]. Sci-fi "metaverse" concepts are slowly becoming reality as products like Fortnite, Minecraft, and Roblox bring immersive social experiences to hundreds of millions of people and blur the lines between games and social networks [24].

Sweeney [24] foundational principles and technologies are applied computing, computing methodologies, and networks and on the other hand computer systems organization, real-time systems, human-centered computing, theory of computing for the metaverse. Siva et al. (2018), to create a virtual environment, use of intelligent avatars and holographic projects, can simulate a real-world classroom scenario.

## IV. CONCLUSION AND RECOMMENDATIONS

It seems that the world of the Metaverse is now clearly coming. This world can harbor both benefits and harms as expected. Just as social media has good and bad sides, the metaverse also has good and bad sides. The hotspot is the usage area. Of course, these issues will be discussed in many scientific studies.

Despite the increasing interest of researchers on the subject of the metaverse, there are few explanatory and comprehensive studies on the subject of a metaverse in the literature. However, it is thought that this situation will increase especially in recent years with the developments in blockchain technology, sensor technology, the advancement of augmented and virtual reality technologies, and the recent statements of South Korea and Facebook founder Zuckerberg.

Facebook is not the only tech company that is invested in exploring a 3D virtual reality where people interact with others using avatars of themselves. Video games such as Roblox and Fortnite are already popular among Gen Z, where they create and interact with each other within their universe [10]. And also they believe this will be the successor to the mobile internet [2]. It can be stated that many different companies may enter the field in this future, enterprises operating in technologically parallel or related sectors are only 3-5 steps ahead of the race at this point, but the sector is fresh and appetizing for entrepreneurs at this point.

Metaverse is not a single product that a company can handle alone. Just like the internet, the metaverse will exist with or without Facebook. It will also not be built overnight. Many of these products will only be fully realized in the next 15-20 years. This situation strengthens the possibility that the subjects will be evaluated from many different aspects in the literature in this development process. This shows the importance of Metaverse and the valuable analysis of the studies carried out in this direction in the literature. It is thought that it will be beneficial for the researchers to compare the data obtained from different databases with the data obtained, and to present a texture analysis on the metaverse periodically, in terms of the development of the field and for the researchers to see the general view.

Facebook is considering recruiting 10,000 new teammates across the EU at Facebook and putting its European presence at the center of plans to help build the metaverse [3] in fact, this situation confronts us with the possibility of many big IT companies entering this field in terms of the metaverse. and it can be seen that it will create new employment areas.

Of course, at this point, it is considered beneficial for higher education institutions to include topics that will be related to the metaverse in some way, such as virtual reality, augmented reality, simulation technology, blockchain technology. In this way, higher education institutions will be able to support and meet the demands of the sector in terms of trained human resources that the sector needs.

## REFERENCES

[1] Alang, N. (2021). Facebook wants to move to 'the metaverse' - here's what that is, and why you should be worried. Access date: 08/11/2021, https://www.thestar.com/business/opinion/2021/10/23/facebook-wants-to-move-to-the-metaverse-heres-what-that-is-and-why-you-should-be-worried.html

[2] Chayka, K. (2021). Facebook Wants Us to Live in the Metaverse. What does that even mean? Access date: 08/11/2021, https://www.newyorker.com/culture/infinite-scroll/facebook-wants-us-to-live-in-the-metaverse

[3] Clegg, N. (2021). Investing in European Talent to Help Build the Metaverse. Access date: 08/11/2021, https://about.fb.com/news/2021/10/creating-jobs-europe-metaverse/

[4] Eschen, H., Kötter, T., Rodeck, R., Harnisch, M., & Schüppstuhl, T. (2018). Augmented and virtual reality for inspection and maintenance processes in the aviation industry. Procedia manufacturing, 19, 156-163.

[5] Espíndola, D. B., Pereira, C. E., Henriques, R. V., & Botelho, S. S. (2010). Using mixed reality in the visualization of maintenance processes. IFAC Proceedings Volumes, 43(3), 30-35.

[6] Gaubert, J. (2021). Seoul to become the first city to enter the metaverse. What will it look like?. Access date 10/11/2021, https://www.euronews.com/next/2021/11/10/seoul-to-become-the-first-city-to-enter-the-metaverse-what-will-it-look-like

[7] Harapan, H., Itoh, N., Yufika, A., Winardi, W., Keam, S., Te, H., ... & Mudatsir, M. (2020). Coronavirus disease 2019 (COVID-19): A literature review. Journal of infection and public health, 13(5), 667-673.

[8] Hardawar, D. (2021). Facebook says it doesn't want to own the metaverse, just jumpstart it. Access date: 08/11/2021, https://www.engadget.com/facebook-connect-metaverse-zuckerberg-171507437.html

[9] Joshua, J. (2017). Information Bodies: Computational Anxiety in Neal Stephenson's Snow Crash. Interdisciplinary Literary Studies, 19(1), 17-47.

[10] Kim, S. (2021). South Korea's Approach to the Metaverse. Access date: 09/11/2021,

[11] Lasry, A., Kidder, D., Hast, M., Poovey, J., Sunshine, G., Winglee, K., ... & Team, R. (2020). Timing of community mitigation and changes in reported COVID-19 and community mobility—four US metropolitan areas, February 26–April 1, 2020. Morbidity and Mortality Weekly Report, 69(15), 451.

[12] Masters, N. B., Shih, S. F., Bukoff, A., Akel, K. B., Kobayashi, L. C., Miller, A. L., ... & Wagner, A. L. (2020). Social distancing in response to the novel coronavirus (COVID-19) in the United States. PloS one, 15(9), e0239025. https://doi.org/10.1371/journal.pone.0239025

[13] Nesbo, E. (2021). The Metaverse vs. Virtual Reality: 6 Key Differences. Access date: 08/11/2021, https://www.makeuseof.com/metaverse-vs-virtual-reality/

[14] Roblox, (2021). What is Roblox? Access date: 09/11/2021, https://corp.roblox.com/

[15] Rotolo, D., Rafols, I., Hopkins, M. M., & Leydesdorff, L. (2017). Strategic intelligence on emerging technologies: Scientometric overlay mapping. Journal of the Association for Information Science and Technology, 68(1), 214-233.

[16] Schlemmer, E. (Ed.). (2014). Learning in Metaverses: Co-Existing in Real Virtuality: Co-Existing in Real Virtuality. Brazil: IGI Global.

[17] Schmidt, J. (2007). Knowledge politics of interdisciplinarity. Specifying the type of interdisciplinarity in the NSF's NBIC scenario. Innovation: The European Journal of Social Science Research, 20(4), 313-328.

*Journal of Metaverse*

*Damar M.*

[18] Schwald, B., & Laval, B. (2003). An Augmented Reality System for Training and Assistance to Maintenance in the Industrial Context. Journal of WSCG, 11, 1-3.

[19] Shotton, D., Portwin, K., Klyne, G., & Miles, A. (2009). Adventures in semantic publishing: Exemplar semantic enhancements of a research article. PLoS Computational Biology, 5(4), e1000361. doi:10.1371/journal.pcbi.1000361

[20] Silva, H., Resende, R., & Breternitz, M. (2018). Mixed reality application to support infrastructure maintenance. In 2018 International Young Engineers Forum (YEF-ECE) Costa da Caparica, Portugal, 4 May 2018 (pp. 50-54). IEEE.

[21] Siyaev, A., & Jo, G. S. (2021). Towards Aircraft Maintenance Metaverse Using Speech Interactions with Virtual Objects in Mixed Reality. Sensors, 21(6), 1-21.

[22] Sparkes, M. (2021). What is a metaverse. New Scientist, 251(3348),1-18

[23] Squires, C. (2021). Seoul will be the first city government to join the metaverse. Access date: 25/11/2021, https://qz.com/2086353/seoul-is-developing-a-metaverse-government-platform/

[24] Sweeney, T. (2019). Foundational Principles & Technologies for the Meta-verse. InProceedings of SIGGRAPH '19 Talks.ACM, New York, NY, USA,1 page. https://doi.org/10.1145/3306307.3339844

[25] Yonhap News Agency (2021). Seoul to offer new concept administrative services via metaverse platform, Access date: 25/11/2021.
http://www.koreaherald.com/view.php?ud=20211103000692

[26] Van Eck, N. J., & Waltman, L. (2010). Software survey: VOSviewer, a computer program for bibliometric mapping. scientometrics, 84(2), 523-538.

[27] Waltman, L., Van Eck, N. J., & Noyons, E. C. (2010). A unified approach to mapping and clustering of bibliometric networks. Journal of informetrics, 4(4), 629-635.